# Superconducting gap structure and pinning in disordered MgB$_2$ films

Y.Bugoslavsky, Y.Miyoshi, G.K.Perkins, A.D.Caplin, L.F.Cohen
*Imperial College London*
H.Y.Zhai, H.M.Christen
*Oak Ridge National Laboratory, Oak Ridge, Tennessee 37931-6056*
A.V.Pogrebnyakov, X.X.Xi,
*The Pennsylvania State University, University Park, PA*
O.V.Dolgov
*Max-Planck-Institut für Festkörperforschung, Stuttgart*

## *Abstract*

We have performed a comparative study of two thin films of magnesium diboride (MgB$_2$) grown by different techniques. The critical current density at different temperatures and magnetic fields was evaluated from magnetisation curves, the structure of superconducting order parameter was obtained from point-contact spectroscopy, and the scattering rates were evaluated by fitting the temperature dependent normal-state resistivity to the two-band model. The films have similar critical temperatures close to 39 K, but the upper critical fields were different by a factor of 2 (5.2T and 2.5 T at 20 K). We have found that the film with higher H$_{c2}$ also had stronger scattering in the σ band and smaller value of the superconducting gap in this band. As the scattering in σ band is primarily due to the defects in boron plane, our results are consistent with the assumption that disordering the boron planes leads to enhanced H$_{c2}$ and better pinning properties in magnetic field.

## *Introduction*

Since the discovery of superconductivity in magnesium diboride (MgB$_2$) this compound has not only become a fascinating scientist's toy, but also promises unrivalled applications. There has been a very rapid progress in understanding the band structure of MgB$_2$ and the double-gap nature of superconductivity, in terms of both theory and experiment. Theoretical guidance that MgB$_2$ possesses two distinct superconducting gaps associated with the two disconnected sheets of the Fermi surface [1] was very soon found to be valid by reproducible measurements using tunnelling spectroscopy [2], point-contact spectroscopy [3], specific heat [4] and de Haas – van Alphen effect [5]. Results of the double-band model were corroborated by treatment of the fully-anisotropic Eliashberg equations [6], which yielded, as an important benefit, detailed variation of the gap values with the direction of the wave vector k. The different dimensionalities of the two order parameters (3D gap in the π band and 2D gap in the σ band) and their behaviour in magnetic field have been also studied in detail [7], and in spite of remaining open questions and some differences in interpretation, there is a broad agreement as for the validity of the multiple-band description of MgB$_2$.

The prerequisite for existence of two distinct superconducting order parameters is weak inter-band scattering. This scattering channel is largely suppressed by the orthogonality of the wave functions in σ and π bands [8]. Therefore the two order parameters can in principle respond differently to disorder, and to describe the resistivity behaviour of MgB$_2$, a model was developed that takes into account different intra-band scattering rates, γ$_π$ and γ$_σ$ [8]. The band disparity adds to the complexity of the superconducting state in MgB$_2$, and at the same time provides additional routes for understanding and possibly fine tuning the disorder in this material.

Disordering a superconductor is one of the ways to improve the pinning properties, and as a result, the critical current density, which is a figure of merit for practical applications. In this respect, considerable progress has been achieved, particularly in oxygen-alloyed thin films [9], proton-irradiated powders [10] and bulk materials with nanometre-size inclusions [11,12]. Recent theoretical calculations of the upper-critical field in a double-gap superconductor [13] show that MgB$_2$ has the prospect of becoming a strong competitor to the superconductors presently used in industrial applications – provided there are viable routes of enhancing both the upper critical field and the in-field critical current.

The progress in producing materials with high critical currents and large critical fields is linked to understanding the effect of various kinds of disorder on the electronic and superconducting properties. In this work we study two differently-grown thin films of MgB$_2$ and correlate their performance in applied magnetic field (critical current density and the value of upper critical field) with the details of the double-gap structure and scattering in the π and σ bands.

## Samples

Film 1 was grown at Oak Ridge National Laboratory by first depositing pure boron on a sapphire substrate using electron-beam evaporation, with a subsequent anneal in a Mg-rich atmosphere. The details of film preparation are described elsewhere [14]. This is the same film that has been studied in our earlier works [15,16]. Film 2 was grown at Pennsylvania State University using Hybrid Physical-Chemical Vapour Deposition [17] also on a sapphire substrate; this film was subject of our recent work [18]. This film has higher resistivity than more recent state-of-the-art HPCVD films, which are very clean [19]. However, as far as the film morphology is concerned, our Film 2 is far superior to the film presented in [20] (in their Fig 1).

It is important for further analysis that both films were dense, with coverage close to 100%. This is illustrated by the SEM image of Film 1 (Fig 1), and the AFM image of Film 2 (Fig. 2). Film 1 is granular and there are small voids visible; the size and the number of voids are small enough not to impede the current flow significantly. In the previous paper [16] we have performed the length-scale analysis of the magnetically-induced supercurrents. The conclusion was that the length scale corresponded to the sample dimensions and was not dependent on magnetic field, up to the irreversibility field. This implies that the film remains globally connected with the supercurrent flow unimpeded by the grain contacts.

Several samples were cut out of the original Film 1. The critical temperature was found to vary slightly between these samples, in the range 37.2 to 38.7 K, as measured by the onset of the diamagnetic screening. The width of the resistive transition in each sample was about 0.5 K (10% - 90% of the normal-state resistance). In Film 2 there was no noticeable variation of $T_c$, but the transition width was slightly larger, $\Delta T_c = 0.7$ K with the mid-point transition at $T_{c\,mid\text{-}point} = 39.0$ K.

## Techniques

Irreversible magnetisation loops were obtained using an Oxford Instruments vibrating sample magnetometer; magnetic fields up to 8 Tesla were applied perpendicular to the films. The critical current density, $J_c$, was evaluated using the relevant formula of the Bean model [21].

The resistance was measured using the standard 4-contact method, in a thin-strip geometry for Film 1, and using van der Pauw method [22] on the square-shaped Film 2. The conversion of the measured resistance into resistivity is subject to a rather large uncertainty, particularly in Film 1, which has significant variation of the thickness. The values of resistivity may also be affected by incomplete connectivity within the films.

Point-contact spectroscopy was performed on junctions created by pressing sharpened gold tips to the films; the tip was always perpendicular to the film surface. In both films double-gap spectra were observed when the background junction resistance was made small (<20 Ohm). Higher-resistance spectra showed only single $\pi$ gap. The spectra were analysed using the double-gap Blonder-Tinkham-Klapwijk (BTK) model [23]. In fitting the experimental data we used five parameters: two gaps ($\Delta_\sigma$ and $\Delta_\pi$), dimensionless barrier Z (the same for the two gaps), relative weight of the two contributions, and generic spectral broadening (the same for the two gaps). The latter included the thermal contribution, but there was a non-thermal part as well, which represents in cumulative form different sources of scattering. We used throughout ballistic-limit formulae for Andreev reflection, although some of the spectra could be fitted as well with the diffusive model [24]. Where comparison was possible, the difference between the two regimes was mainly in the inferred values of the barrier parameter Z. The diffusive regime predicts a deep minimum at zero bias even for Z=0, as opposed to a wide peak in the ballistic case. Non-zero Z is necessary for the minimum to appear in the ballistic regime. Therefore attempting to fit the same data with two models results in the diffusive fit producing significantly smaller values of Z than the ballistic fit. However the choice of the regime did not affect other parameters, most importantly the gap values. The ballistic fit worked well with nearly 100% of experimental double-gap spectra, and the uncertainty of the gap values is estimated to be not worse than 0.1 meV.

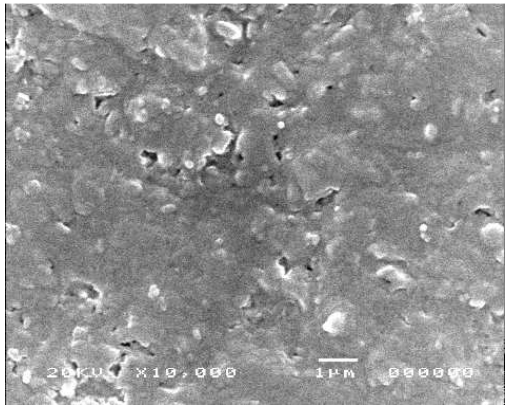

Fig 1. SEM image of the surface of Film 1 shows nearly complete coverage, but granular structure and a few inter-grain voids are visible.

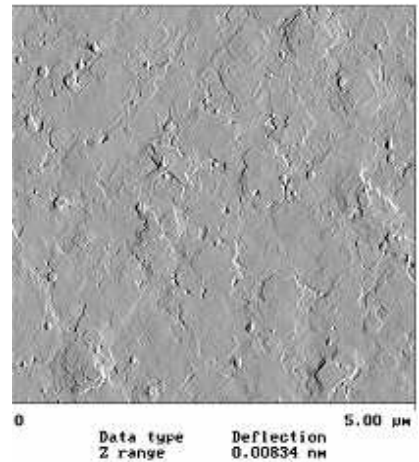

Fig 2. Deflection-mode AFM image of Film 2. This film is porous-free and continuous. Spiral growth columns are clearly visible, indicative of epitaxial growth.

## *Results and discussion.*

It has been observed that the enhancement of pinning in $MgB_2$ is often associated with a rather substantial decrease in $T_c$ [9, 16]. This is not the case with our films: their behaviour in magnetic field is dramatically different in spite of very similar $T_c$'s. The critical current densities derived from the width of the magnetisation loops are shown in Fig 2. As was confirmed by the length scale analysis, the films remained connected superconductors in applied field, and the decrease of magnetisation reflects a genuine decrease of $J_c$, not magnetic fragmentation of the sample due to weak links or grain boundaries. Importantly, this is valid for randomly-oriented Film 1. We can assume therefore that, with individual grains anisotropic, the disappearance of global irreversible magnetisation indicates the smallest orientation-dependent value of the $H_{c2}$, which corresponds to the field parallel to the grain's c axis. The epitaxial film was measured with H||c axis as well, which makes the comparison of the critical fields in the two films meaningful. As with the calculation of resistivity, the evaluation of $J_c$ from the magnetic moment is subject to the uncertainty in sample geometry, which may account for the lower value of the $J_c$ at zero field in Film 1. In Film 2 our estimate of $J_c$ at 4.2 K, zero field is $J_c = 31$ $MA/cm^2$, which is in excellent agreement with the $J_c$ of the best film measured in [20] by the transport method, which was 34 $MA/cm^2$ at 5 K. The geometrical issues do not affect the consideration of $H_{irr}$ and $H_{c2}$ (apart from the discussed possibility of magnetic fragmentation). The irreversibility field, $H_{irr}$, in films nearly coincides with the upper critical field $H_{c2}$. We have checked that the faster depression of $J_c$ by magnetic field in film 2 is indeed due to its lower $H_{c2}$. We have measured resistive transitions in applied field, and from these plotted the diagrams of $H_{c2}(T)$ (taken to be corresponding to 90% of the normal-state resistance, $R_n$) and $H_{irr}(T)$ (10% of $R_n$). At 20 K the values of $H_{irr}$ and $H_{c2}$ were 2.5 T and 2.9 T, respectively. As shown in Fig 2, $H_{irr}$ determined from the resistivity measurements is in perfect agreement with that obtained from magnetic $J_c$ with the criterion set at 1 kA/cm$^2$. It is also clear from that graph that the $H_{c2}$ in Film 1 is almost twice as high than in Film2. In both films the temperature dependence of $H_{irr}$ is closely linear (Inset, Fig 2), apart from the curvature close to $T_c$ in Film 1.

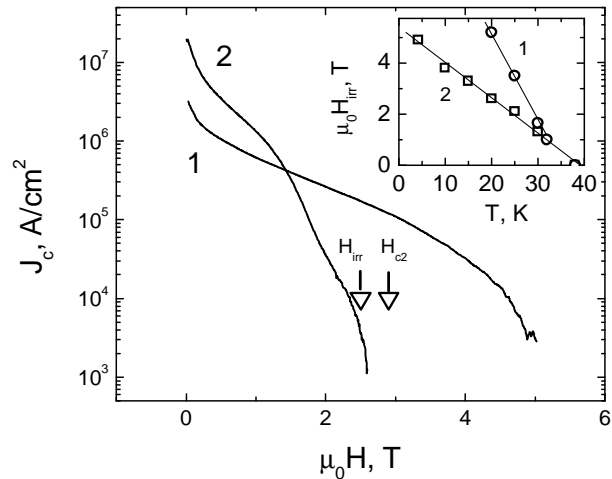

Fig 3. Critical current density vs applied magnetic field at 20 K for the two films; the field applied perpendicular to the films. The arrows indicates the irreversibility field $H_{irr}$ and the upper critical field $H_{c2}$ of Film 2, as obtained from the resistivity measurements. Inset: the irreversibility field vs temperature.

In the following we attempt to establish the microscopic reason behind such a large difference in $H_{c2}$. Since the behaviour of $H_{c2}$ in MgB$_2$ is strongly dependent on the multiple-band electronic structure [13], we have measured point-contact spectra of the films to obtain details of the double-gap superconducting state.

In both films the point-contact spectra showed clear signatures of two gaps in both films (Fig.4). The spectra are consistent with the results obtained by other groups on various samples of MgB$_2$ [7,25,26]. It has been proven (by comparison with the heat capacity measurements [4] and de Haas – van Alphen effect [5]) and is now accepted that the gap structure measured by surface-sensitive methods reflects bulk properties of MgB$_2$. Unlike tunnelling spectroscopy, the point contact is not a directional probe. With the tip tightly pressed to the sample, electrons are injected into a broad cone, so probing a wide range of k-space simultaneously. A number of junctions were measured on each film and the values of the gaps inferred from the fitting procedure were found to vary from junction to junction. Statistical distributions of the measured gaps are shown in Fig 5. It is natural to compare these results with the theoretical intrinsic distribution calculated by Choi et al. [6]. In the limit of negligibly small scattering, the gap values in π and σ bands are dependent on the direction of the wave vector, so that probing all the direction with equal weights produces the distribution shown in Fig 5 (c). The comparison between theory and experiment has to be done with caution though, as scattering in the films is strong and the clean limit does not apply. With this high scattering rates any microscopic variation of the gaps over the Fermi surface should be averaged out, so that the directional dependence should not be observable. Still the measured gaps vary by an amount that is greater than the uncertainty of the fitting. This point is illustrated in Fig 4, which shows that the difference in measured values of $\Delta_\pi$ is by far greater than any fitting-related uncertainty. The exact nature of this variability has yet to be understood. The most likely explanation is that the surface-state gaps are shifted slightly by surface-induced inter-band scattering. To enhance the inter-band process in the bulk requires impurities that break the disparity between σ and π bands. On the other hand, similar symmetry breaking can also occur at the surface. This effect would allow for some degree of variability between the local (at a particular point at the surface) and the bulk gaps. It has to be noted however that the σ-π scattering would tend to merge the gaps. It has been suggested that the merging would occur by simultaneous shift of the σ gap down and the π gap up [27]. Recent doping studies suggest otherwise: the π-σ scattering depresses the σ gap, but leaves the σ gap almost in place [28]. If the former applied, we should expect a correlation between positions of the π and σ gaps – but no such correlation has been observed. Alternatively, the variation of the π gap position would be significantly smaller than the variation of the σ gap – which was neither the case, as the gaps varied by nearly the same amount. On the other hand, if the local values of $\Delta_\sigma$ were indeed affected by the inter-band scattering, the corresponding local variation of $T_c$ would be small and comparable with the width of the superconducting transition.

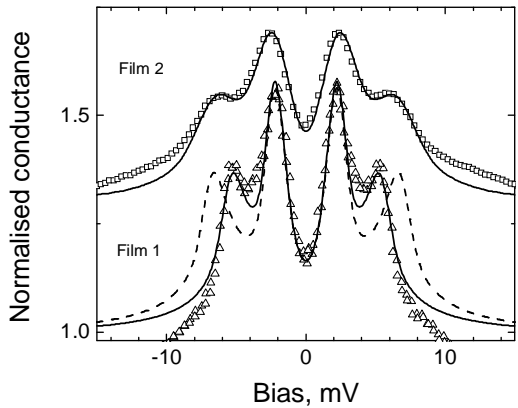

Fig 4. Examples of the point-contact spectra on Film 1 (a) and Film 2 (b) at 4.2 K and in zero magnetic field. Solid lines are best fits to the two-gap BTK model. The fitting is robust as for the values of the gaps, so that the difference in Δ between spectra is greater than the uncertainty in determining the values. The value of $\Delta_\sigma$ is 7.1 and 5.6 meV in the top and bottom graphs, respectively. The dashed line shows how the bottom spectrum would have looked like if $\Delta_\sigma$ was 7.1 meV.

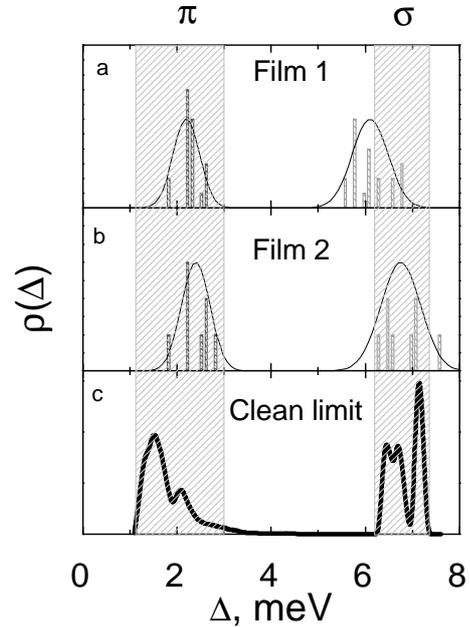

Fig 5. Statistical distribution of the measured gap values compared to the results of clean-limit theory [6]. The distributions of $\Delta_\pi$ are almost identical in the two films, but shifted towards the higher-energy wing of the theoretical profile. In Film 2, the measured values of $\Delta_\sigma$ fall within the theoretical range, whereas in Film 1 the values are statistically lower.

We can however draw positive conclusions from the fact that statistically the gap structures in the two films are different. Indeed, as Fig 5 demonstrates, in Film 2 the histograms for both σ and π gaps lie within the bounds of the theoretical distribution. If we adopt the hypothesis of surface-induced σ-π scattering, the coincidence of the distributions would mean that the scattering is in fact small and only introduces shifts to the gaps that are smaller than the theoretical dispersion, so that the measured values are always within the ranges "allowed" by the clean-limit theory. The situation is different in Film 1. Here the σ gap histogram is shifted to lower energies and partly out of the theoretical range. The value of the shift is significantly larger than the accuracy of the measurement, although comparable with the width of the distribution. The maximum count for σ gap in Film 1 occurs at a value of 5.7 meV, whereas in Film 2 no σ gaps have been seen at energies below 6.2 meV. In contrast to the σ gaps, the distributions of π gaps are almost identical in the two films.

Since the distribution of gap values within each film most probably reflects weak surface-induced π-σ scattering, the consistently smaller σ gap in Film 1 must be a bulk effect. It indicates the presence of pair-breaking scatterers in this film. This assumption agrees with the depressed $T_c$ in this film. On the basis of the two-band model [13], Ferrando et al [29] estimated the relation between suppression of $T_c$ and the inter-band scattering rate, $\gamma_{\pi\sigma}$. From their experimental results, $\delta T_c[K] \approx -2.3\, \gamma_{\pi\sigma}[meV]$. If we identify the depression of the mean value of $\Delta_\sigma$ in Film 1 ($\delta\Delta_\sigma = 0.6$ meV) with the inter-band scattering rate, the corresponding suppression of $T_c$ would be 1.4 K, in agreement with the measured result. We note also that it has been shown in doping experiments [28] that, unlike early theoretical predictions, the inter-band scattering scarcely affects the π gap, so that even in samples with $T_c$ around 20 K it stays close to 2 meV. It is therefore not surprising that in our films there is no visible difference in π gap distributions.

Although the π-σ scattering is enhanced in Film 1, it is still far from being the main scattering channel, so the dominant effect on $H_{c2}$ should come from the intra-band processes. To evaluate those, we turn to the resistivity dependences, ρ(T), as shown in Fig 6 The resistivity at 40 K is noticeably larger than in clean single crystals with comparable $T_c$, indicating significantly stronger scattering in the films. The behaviour of the normal-state resistivity in the two films was different. If one assumes that there are only intrinsic scattering mechanisms, the ρ(T) dependence can be fitted using the two-band model with different intra-band scattering rates [30]. The values estimated from the fit are as follows: $\gamma_\sigma = 71$ meV, $\gamma_\pi = 220$ meV; and $\gamma_\sigma = 27$ meV, $\gamma_\pi = 950$ meV for films 1 and 2, respectively.

We recognise that these numbers may be subject to large uncertainty, as the resistivity may be overestimated due to variation of the thickness across the films, incomplete connectivity between the grains and extrinsic scattering. Nevertheless a qualitative argument can lend support to the general validity of these estimates.

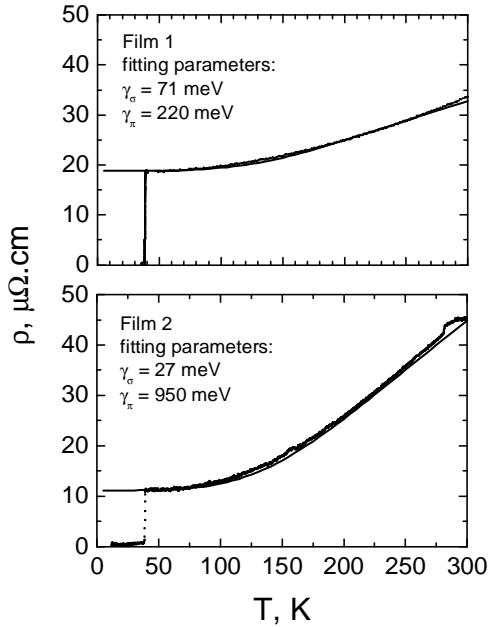

Fig 6. Resistivity vs temperature for Film 1 (a) and Film 2 (b). Solid lines are best fits to the two-band model with the parameter values as indicated.

In the two-band model, the shape of $\rho(T)$ is dependent on the scattering rates $\gamma_\sigma$ and $\gamma_\pi$ in a non-trivial way. It is however safe to say that to have $\rho(40K)$ of the order of 10 $\mu$Ohm cm certainly requires the $\pi$ band to be in the dirty limit – even if the numbers for $\gamma_\pi$ coming from the fitting are overestimated. Following the argument of Mazin et al [8], we can assume that the temperature dependence of $\rho$, and particularly the residual resistivity ratio (RRR) largely reflect the properties of the $\sigma$ band. Qualitatively, a weak temperature dependence (and hence lower RRR) would then indicate relatively large impurity scattering compared to the phonon scattering. Therefore smaller RRR in Film 1 qualitatively agrees with the estimated $\gamma_\sigma$ in this film being larger than in Film 2. The overall conclusion is that $\pi$ bands are well into the dirty limit in both films, and the $\sigma$ band is dirtier in Film 1 than in Film 2.

These results can be compared with the dirty-limit theory for the upper critical field in a two-band superconductor [13]. The theory operates with electron diffusivities, which are inversely proportional to scattering rates. The theory predicts that depending on the ratio between electron diffusivities in the two bands, the temperature variation of $H_{c2}$ is different, with positive or negative curvature. At temperatures above approximately $T_c/2$ the dependence is closely linear in all regimes. The linearity can be retained to low T if $D_\pi \approx 0.4\, D_\sigma$, i.e. when the $\pi$ band is about twice as dirty as the $\sigma$ band. If the diffusivities in $\sigma$ and $\pi$ bands are different, the $H_{c2}$ at temperatures close to $T_c$ is determined by the band with maximum diffusivity, i.e., minimum scattering rate. The experimental results appear to agree with the conclusions of the theory. The strong $\pi$-band scattering in both films brings them into the regime where the slope of $H_{c2}(T)$ is determined by the $\sigma$ band. It is likely that the values of $\gamma_\pi$ are over-estimated and indeed the difference between $\gamma_\pi$ and $\gamma_\sigma$ is not so large as to cause positive curvature of the $H_{c2}(T)$ dependence. Hence from the theory we should expect that the slope of $H_{c2}(T)$ should be larger in the film with stronger $\sigma$-scattering, which is our Film 1 – exactly as we have observed.

Different dimensionalities of the bands and the associated symmetries of the wave functions allow one to attribute scattering channels to certain classes of structural defects. Enhancement of $\gamma_s$ requires substitutions in the boron plane (for example, by carbon or oxygen). The inter-band scattering occurs when there are out-of-plane distortions in the boron sheets, which may be caused by substitutions (but not vacancies) in the magnesium layers [31]. The fact that both these channels appear enhanced in our Film 1 suggests that both magnesium and boron subsystems contain considerable disorder, whereas Film 2 contains more ordered boron planes, and as a consequence, weaker scattering in the $\sigma$ band and lower upper critical field.

## *Conclusions*

We have studied two different $MgB_2$ films with similarly high $T_c$'s but with upper critical fields different by a factor of 2. The analysis of point-contact spectra and the normal-state resistivity shows that the enhancement of $H_{c2}$ in Film 1 is associated with stronger scattering in $\sigma$ band compared to that in Film 1. There is also a detectable enhancement of inter-band scattering in Film 1, although it remains much smaller than the intra-band scattering rates. This suggests that Film 1 contains defects in the boron plane ($\sigma$ scattering) as well as pair-breaking inter-band scatterers in the magnesium plane.

## *Acknowledgements*

We are grateful to I.I.Mazin, J.Kortus and A.Gurevich for useful discussions. This work was supported by the UK Engineering and Physical Sciences Research Council. Work at Penn State was supported in part by ONR under grant No. N00014-00-1-0294